# Guarded by Gamora: How Access Control Balances Out Waiting Times in Transport Systems

Pasquale Grippa, Evşen Yanmaz, Paul Ladinig, and Christian Bettstetter

*Abstract*—A transport system with passengers traveling between stations in periodically arriving cabins is considered. We propose and evaluate an access control algorithm that dynamically limits the number of passengers who are allowed to board the current cabin. Simulation of a ski lift using empirical passenger data suggests that such access control can balance out the average waiting times at different stations. The algorithm works well with estimated values of the passengers' arrival and de-boarding rates.

## I. INTRODUCTION

Public transport systems have a broad range of design options and control mechanisms for improving the passenger throughput and travel experience. For example, they can be either schedule-based or on demand, fixed or adaptive to passenger load, and may or may not provide feedback to passengers and implement travel redirection. Regardless of the requirements and constraints, a passenger flow control can be an effective and economical method of reducing congestion and deadlock, while guaranteeing reliability and safety.

To contribute to this domain, we propose and evaluate an *access control algorithm* for implementation in the boarding areas of ski resorts, where customers use fixed-capacity, fixed-speed cabins between stations. Based on real-time knowledge of passenger arrival and queuing conditions at all the stations, the algorithm adaptively reports as to how many passengers are allowed to enter the next cabin. Metaphorically speaking, it acts as a *guard* for the cabins. The objective is to automatically improve passenger comfort and fairness by having waiting times in the same order of magnitude at all stations. The concept is evaluated by simulation using real passenger data obtained from the Austrian ski resort Bad Gastein, where it was tested by an industrial partner.

The algorithm computes the maximum number of passengers who can enter the next cabin based on the measured arrival rates, so that the stability thresholds are equalized. To this end, it utilizes the queuing model and analytical expressions derived in [1], in which each station is modeled as a queue with Poisson arrivals and bulk services with deterministic service time. There are many studies on queuing systems with Poisson arrivals, where the queue is served in batches with a given maximum size and a random, independently-distributed time between consecutive services

P. Grippa and C. Bettstetter are with the Institute of Networked and Embedded Systems, Alpen-Adria-Universität Klagenfurt, Austria. E. Yanmaz, P. Ladinig, and C. Bettstetter are with Lakeside Labs GmbH, Klagenfurt, Austria. Email: pasquale.grippa@aau.at.

(see [2]–[6]). However, in those papers, the server is assumed to serve a fixed batch size. Extensions exist such that queue length expressions can be derived for lifts, where part of the capacity is already in use [7]. In our model, we considered a generic distribution of the capacity. Based on the arrival rate and capacity at a station, we derived stochastic properties of the waiting time and queue length [1].

The performance of the access control algorithm is compared to that of no control and static control. The latter reserves a fixed number of seats at the ground station for use at succeeding stations. Our results show that the algorithm achieves the best balance of waiting times at the stations. In real scenarios, the parameters of the stochastic processes regulating passengers' arrival and leaving are not known and have to be estimated. According to our results, the algorithm is robust with respect to the estimation of these parameters.

The paper is organized as follows. Section II discusses related work. Section III introduces the system model and notation. Section IV explains the algorithm. Section V presents and discusses the results.

## II. RELATED WORK

The access control of passengers to ski lifts or similar systems has not been investigated in the scientific literature. Although the problem is related to ramp metering control on highways, which is well investigated, there are some important differences between the two problems. An overview and literature review on ramp metering is presented in [8] and [9], and some algorithms are evaluated in [10]. A common solution for ramp metering is ALINEA [11], which seeks to keep the highway occupancy to a desired value and contains a feedback law developed using classical control theory. It regulates a single ramp but does not consider queue length and waiting time on the ramp. Other approaches consider the transport network and seek to minimize the total time spent in the system including the waiting time on the ramps. These approaches use non-linear programming [12], neuro-fuzzy algorithms [13], and model-predictive control [14].

The main differences between ramp metering and our problem are the system model and the performance metric. Highway traffic is usually modeled [14] using a set of equations relating to traffic density (vehicles per unit space), vehicle mean speed, traffic flow (vehicles per unit time), and other values. Equations for modeling the ramps include traffic demand (vehicles per unit time), queue length (vehicles), and metering rate (control variable). These quantities are not modeled stochastically, i.e., are not associated with probability distributions. Another difference is that it is usually possible to measure the traffic flow on highways but, in our experience, not always on ski lifts. For this reason, we do not track cabin occupancy but have based our algorithm on a stochastic model [1]: The free places in cabins are represented with a random variable (capacity), and the access control changes the distribution of this variable.

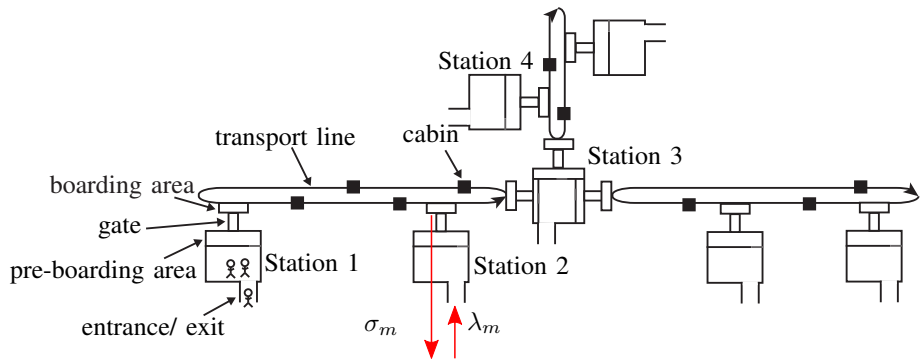

Fig. 1: Transport system model.

The performance of ramp metering is usually evaluated by the total time spent in the system, i.e., the sum of the periods on the highway and ramp. Highway congestion must be avoided to keep this time low. In contrast, in our problem, the number of passengers riding in the cabins does not influence the cabin speed, i.e., there is no congestion on the transport line. For this reason, we have not focused on the total time spent in the system but on achieving a fair waiting time at all the stations. To be precise, we are not interested in having exactly the same waiting time at all stations, as this would not translate into a higher quality of service for the passengers. Instead, we want to have waiting times of the same order of magnitude, to avoid passengers at one station waiting for hours and those at other stations waiting for only minutes.

Many papers on transit networks focus on the transit assignment problem, in addition to passenger flow modeling (see [15]–[17]). However, the main goal in such networks is to find the best routes for passengers (e.g., with minimum travel time, most comfort, least cost), given the passenger load, preferences, and demands. Furthermore, there typically are many alternative paths to certain destinations via different modes of transportation that run on schedule [18], whereas in our problem, the passenger behavior is not directly considered, and the access decisions are made by the transport system itself. In particular, the system being studied here has periodically arriving (always available) cabins and the goal of assigning the passengers to cabins in a seamless way throughout the transport line via an access control algorithm.

## III. SYSTEM MODEL AND PRELIMINARIES

A transport network as shown in Figure 1 is composed of stations connected by transport lines. Cabins move along the lines to carry passengers from one station to the next. Some stations act as junctions, enabling passengers to change lines. Stations on the same line share cabins.

The system operator can reserve cabin seats at certain stations, with the goal of avoiding passenger congestion at other stations. Such a reservation mechanism can be implemented using a boarding gate that displays the number of passengers allowed to enter an incoming cabin [19]. We propose an algorithm, called *Gamora*, for such access control; it computes the number of passengers allowed to enter a certain cabin and adapts this output to the passenger load.

Some of the ideas behind Gamora arose from our modeling of the transport system as a queuing network [1] and the associated analysis of stability and waiting times in stationary conditions for a single transport line with Poisson arrivals of passengers. At the core of Gamora is, however, the equalization of the scaled stability thresholds, which are computed in closed-form for general arrivals. The algorithm can therefore be applied for all lines of the system.

We indicate stochastic variables and processes with capital Latin letters, system parameters with Greek letters, and vectors in boldface. At station $m$, passengers arrive with rate $\lambda_m$. Cabins with $\gamma$ seats (cabin size) move along the line stopping at stations at constant time intervals $\beta$. Whenever a cabin stops at station $m$, each passenger in the cabin leaves with probability $\sigma_m$, making a seat available for the passengers waiting at the station. Before the $n$th service, the boarding buffer is filled with at most $\eta_{m,n}$ passengers, which is computed by the algorithm. The passengers waiting at station $m$ see a number of free seats, which is a realization of the stochastic process $C_{m,n} = \min[\eta_{m,n}, \gamma - S_{m,n}]$, where $S_{m,n}$ is the number of passengers remaining in the cabin after others leave. The number of passengers entering the cabin $T_{m,n}$ depends on both the capacity $C_{m,n}$ and the number of waiting passengers $Q_{m,n}$. One objective is to have waiting times $W_{m,n}$ of the same order of magnitude at all stations. The service index $n$ is omitted for simplicity in the following.

## IV. ALGORITHM

Gamora's task is to compute, for each incoming cabin, the number of passengers allowed to board. As input, it takes the average number of passengers in the cabins arriving at the first station, arrival rate at all stations, number of passengers waiting at all stations, probability of leaving at all stations, interarrival time between cabins, and cabin capacity. It computes the scaled stability thresholds, determines which stations need additional capacity, and returns the maximum number of passengers who can enter the next cabin, given the measured arrival rates, so that the scaled stability thresholds are equalized. The parameters are summarized in Table I. It is important to note that the *average number of passengers waiting at all stations* is an output from the system which is used as an input into the algorithm, according to $\lambda_{m,\text{in}} = Q_m/\beta + \lambda_m$. This is done to create a negative feedback loop, which counteracts variations in the arrival rate: If $\lambda_2$ decreases, $\lambda_{2,\text{in}}$ decreases, $\eta_1$ increases, which over time increases $Q_2$, which in turn increases $\lambda_{2,\text{in}}$.

TABLE I: Gamora Parameters

**Input**

| | |
|---|---|
| $r_0$ | average number of passengers in the cabins arriving at the first station ($r_0 = \text{E}[R_0]$) |
| $\boldsymbol{\lambda}$ | arrival rates at all stations |
| $\boldsymbol{Q}$ | number of passengers waiting at all stations |
| $\boldsymbol{\sigma}$ | probability of leaving at all stations |
| $\beta$ | interarrival time between cabins |
| $\gamma$ | cabin size |

**Output**

| | |
|---|---|
| $\boldsymbol{\eta}$ | maximum accesses per service at all stations |

**Others**

| | |
|---|---|
| $\boldsymbol{\nu}$ | fraction of arrivals at all stations ($\boldsymbol{\nu} = \boldsymbol{\lambda}/\lambda_{\text{tot}}$) |
| $\bar{c}$ | average capacity at all stations |
| $\boldsymbol{\lambda}_{\text{s}}^*$ | scaled stability threshold at all stations |
| $\boldsymbol{T}$ | number of passengers entering one cabin all stations |

Before we present the algorithmic details, let us explain why we use scaled stability thresholds as decision criteria.

### A. Scaled Stability Thresholds and Waiting Time

Consider the transport line connecting the first three stations in the transport network in Figure 1. Passengers can leave their cabins at each station with a certain probability (empirical values are provided) and can transfer to the other line at Station 3. The arrival rate at station $m$ can be expressed as a fraction of the total arrival rate, $\lambda_m = \nu_m \lambda_{\text{tot}}$, and the expected waiting time can be plotted over the total arrival rate (see Figure 2 showing stationary performance).

The scaled stability threshold $\lambda_{\text{s},m}^*$ is the smallest arrival rate (total) such that station $m$ is unstable, i.e., the expected waiting time never reaches a stationary value but steadily increases over time. If the system is neither underutilized nor unstable (trivial conditions), the total arrival rate at which the system operates is near $\lambda_{\text{s},2}^*$, i.e., $W_2 \gg W_1$. As proved in [1], it is impossible to reduce $W_2$ by orders of magnitude without reducing the stability of the whole system. However, it is possible to reduce $W_2$ to an extent that is interesting for a real application. For instance, as shown in Figure 2, $W_2$ is reduced by up to 50% if two seats are always reserved ($\eta_1 = 6$). This is achieved by decreasing $\lambda_{\text{s},1}^*$ as much as possible without crossing $\lambda_{\text{s},2}^*$. Since, in non-stationary conditions, the fraction of arrival rate at the stations $\nu_m$ changes over time, *the scaled stability thresholds must be continuously adapted*.

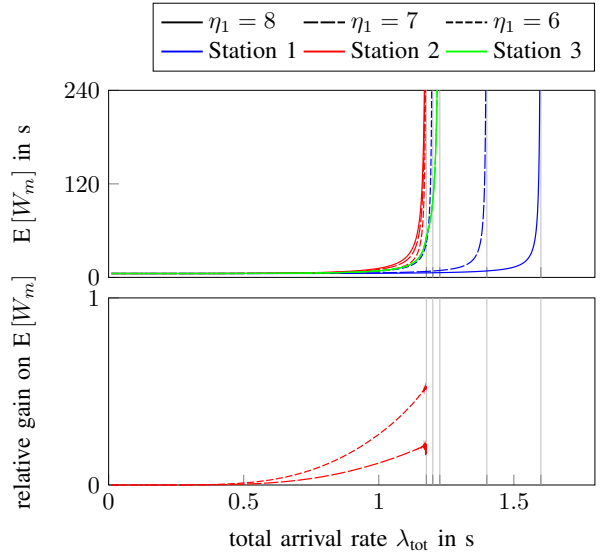

Fig. 2: Relative gain in expected waiting time at Station 2 by reserving one or two seats at Station 1. Parameters: $\beta = 10$, $\gamma = 8$, $r_0 = 0$, $[\nu_1, \nu_2, \nu_3, \nu_4] = [0.5, 0.2, 0.3, 0]$, $[\sigma_1, \sigma_2, \sigma_3, \sigma_4] = [0, 0.04, 0.46, 1]$. Figure taken from [1].

### B. Access Control Algorithm: Gamora

The idea described in the section above is implemented in two steps (see Algorithm 1): First, the transport line is divided into blocks such that the scaled stability threshold of the last station in each block is the smallest within the block and is smaller than all succeeding scaled stability thresholds (*while* loop). Second, the scaled stability thresholds in each block are adjusted (*for* loop).

In Algorithm 1, $\boldsymbol{b}[1]$ is always zero, while $\boldsymbol{b}[i]$ is the index of the last station included in the block $i$. For instance, if $\boldsymbol{b} = [0, 3, 5, 8]$, the system is divided into three blocks: 1 to 3, 4 to 5, and 6 to 8.

The function STABILITY($r_0, \boldsymbol{\nu}, \boldsymbol{\sigma}, \beta, \gamma$) computes the expression (see [1])

$$\boldsymbol{\lambda}_{\text{s}}^*[m] = \frac{\gamma - r_0 \prod_{i=1}^{m}(1-\sigma_i)}{\sum_{j=1}^{m} \nu_j \beta \prod_{i=j+1}^{m}(1-\sigma_i)}. \quad (1)$$

This expression is only exact for the smallest scaled stability threshold in the system. Since we have a division into blocks, the expression is always used correctly.

The function GAMORAB($r_0, \boldsymbol{\nu}, \boldsymbol{\sigma}, \beta, \gamma, \boldsymbol{\lambda}_{\text{s}}^*$) in Algorithm 2 controls the maximum number of accesses per service $\boldsymbol{\eta}[k_1+1, k_2]$ within the block. It evaluates the stability of the stations with respect to the last station in the block, starting from $\boldsymbol{\eta}[m] = 1$ and increasing it until the station is more stable than the last one. This evaluation requires the computation of the expected capacity.

The function CAPACITY($\bar{r}_0, \boldsymbol{\nu}, \boldsymbol{\sigma}, \beta, \gamma, \boldsymbol{\lambda}_{\text{s}}^*, \boldsymbol{\eta}$) iteratively applies the following equation (see [1]) from the first to the

## Algorithm 1 Access Control Line

```
1: function GAMORA (r_0, λ_in, σ, β, γ)
2:     η ← [γ, γ, ..., γ]
3:     ν ← λ_in/TOTAL(λ_in)
4:     λ_s ← [ ]
5:     b ← [0]
6:     k ← 1
7:     while k ≤ Length[λ_in] do
8:         if k=1 then r_in ← r_0
9:         else r_in ← γ
10:        end if
11:        λ_tmp ← STABILITY(r_in, ν[k, end], σ[k, end], β, γ)
12:        λ_s ←append min[λ_tmp]
13:        b ←append argmin[λ_tmp] + k − 1
14:        k ← b[end] + 1
15:    end while
16:    for i ∈ [1, Length[b] − 1] do
17:        if i=1 then r_in ← r_0
18:        else r_in ← γ
19:        end if
20:        k_1 = b[i]
21:        k_2 = b[i+1]
22:        ν' ← ν[k_1+1, k_2]
23:        σ' ← σ[k_1+1, k_2]
24:        η[k_1+1, k_2] ← GAMORAB(r_in, ν', σ', β, γ, λ_s[i])
25:    end for
26:    return η
27: end function
```

## Algorithm 2 Access Control Block

```
1: function GAMORAB(r_0, ν, σ, β, γ, λ_s*)
2:     η ← [γ, γ, ..., γ]
3:     for m ← 1, Length[ν] do
4:         η[m] ← 1
5:         c̄ ← CAPACITY(r̄_0, ν, σ, β, γ, λ_s*, η)
6:         while c̄[m] < ν[m]λ_s*β and η[m] < γ do
7:             η[m] + +
8:             c̄ ← CAPACITY(r_0, ν, σ, β, γ, λ_tot, η)
9:         end while
10:    end for
11:    return η
12: end function
```

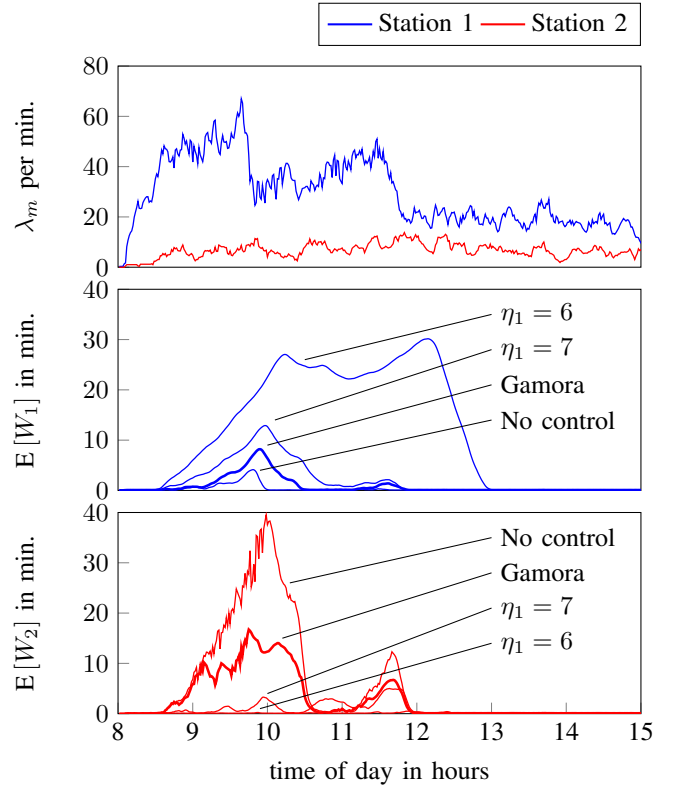

Fig. 3: Average waiting time over the day for arrival rates collected at the Bad Gastein ski resort. The Gamora algorithm best balances out the waiting times at the two stations.

last station of the block:

$$\mathrm{E}[C_m] = \min\left[\eta_m, \gamma - r_0 \prod_{i=1}^{m}(1-\sigma_i) - \sum_{j=1}^{m-1} \mathrm{E}[T_j] \prod_{i=j+1}^{m} 1 - \sigma_i\right]$$

with $\mathrm{E}[T_m] = \min[\nu_m \lambda_{\mathrm{tot}} \beta, \mathrm{E}[C_m]]$.

## V. RESULTS AND DISCUSSION

The algorithm performance is evaluated using a discrete-event simulator. Passenger arrivals are generated by a Poisson process with a rate that varies over time, according to data collected from the Bad Gastein ski resort. All the simulation results are averaged over 35 simulation runs.

### A. Expected Waiting Times

We first analyze the impact of access control on waiting times. The top plot in Figure 3 shows the arrival rates over a day from 8:00 to 15:00. At the beginning of the day, all the passengers queue at the entrance to the ski resort (Station 1). A second peak occurs at noon, due to half-day skiers. The arrival rate at Station 2 is very low because few people ski back to this station. The second and third plots show the average waiting times at Stations 1 and 2, respectively. Access is studied without control, with static control, and with Gamora. The static control reserves one seat ($\eta_1 = 7$) or two seats ($\eta_1 = 6$) at Station 1 for use at Station 2.

In the first two hours, the arrival rate at Station 1 is so high that the system becomes unstable. Thus, without control, the cabins arriving at Station 2 are almost always full, which means that passengers experience long waiting times there. This waiting time can be shortened drastically if the system operator applies static control by reserving one or two seats at Station 1, but this greatly increases the waiting time at Station 1, since the waiting passengers are constantly denied boarding. Using Gamora instead of static control yields much more balanced waiting times across the stations.

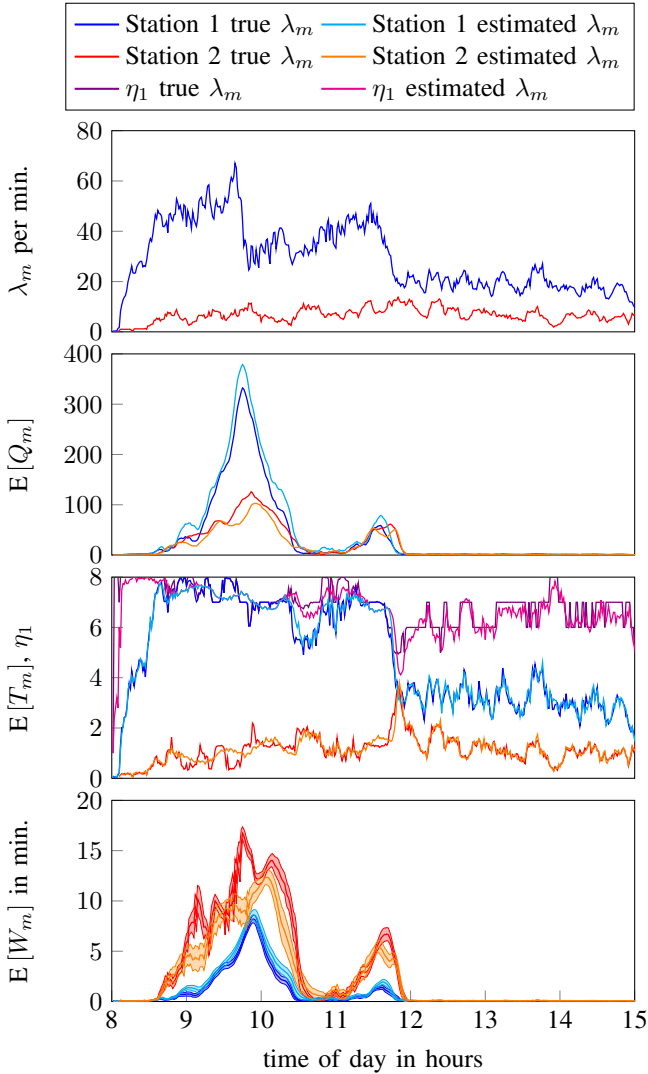

Fig. 4: Performance for true and estimated $\lambda_m$. The algorithm is robust regarding the estimation of this parameter. Confidence intervals are computed with a confidence of 95%.

## B. Impact of Parameter Estimation

In a real system, neither the arrival rates $\lambda_m$ nor the probabilities of leaving $\sigma_m$ are known and must be estimated during operation. We thus investigate Gamora's robustness to errors in the estimations.

Figure 4 compares the performance when using true versus estimated arrival rates. The arrival rate is estimated by counting the number of passengers arriving over $\beta$ and normalizing to $\beta$. Then, for the estimated case, the input of the algorithm $\lambda_{m,\text{in}} = Q_m/\beta + \lambda_m$ is averaged over the last 20 minutes. The first plot shows the true arrival rates at each station. The second and third plots show that estimating these rates has little influence on the queue length, algorithm output, and number of passengers boarding. The last plot shows that average waiting times do not suffer because of an estimate being used.

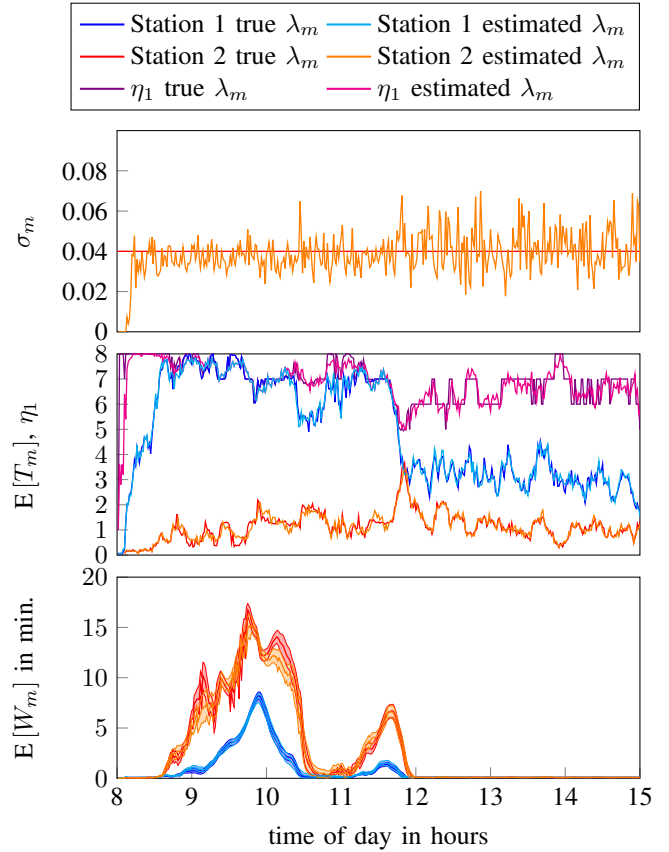

Fig. 5: Performance for true and estimated $\sigma_m$. The algorithm is robust regarding the estimation of this parameter. Confidence intervals are computed with a confidence of 95%.

Figure 5 shows that Gamora is also robust if the leaving probability $\sigma_m$ is estimated. The easiest way to estimate $\sigma_m$ would be to count the number of passengers leaving the cabin at each service. Typically, counting sensors for individual cabins are not installed; it is more common to have sensors at the entrance and exit to the entire station. We thus estimate $\sigma_2$ as the number of passengers leaving Station 2 divided by the number of passengers entering at Station 1 over four minutes (these two values are sampled taking into account the transport delay). Estimating $\sigma_m$ results in only minor changes in the average waiting time of the system.

ACKNOWLEDGMENTS

This work was supported by Universität Klagenfurt and Lakeside Labs GmbH, Klagenfurt, Austria, and funding from the European Regional Development Fund and the Carinthian Economic Promotion Fund (KWF) under grant 20214 a. 3520/27678/39824. The authors would like to thank Jorge Clemente and Andreas Kerschbaumer (SKIDATA) for providing the empirical passenger arrival data and Udo Schilcher (Lakeside Labs) for processing it.